\documentclass[11pt,twoside]{article}

\usepackage{asp2006}
\usepackage{epsf}
\usepackage{psfig}
\usepackage{lscape}

\begin{document}
\title{Dynamical Age vs Spectral Age of the Lobes of Selected Giant Radio Sources (GRSs)}   
\author{J. Machalski,$^1$ M. Jamrozy,$^1$ and D.\ J. Saikia$^2$} 
\affil{$^1$Astronomical Observatory, Jagiellonian University, Krakow, Poland  \\  
$^2$National Centre for Radio Astrophysics, TIFR, Pune, India}

\begin{abstract} 
Dynamical ages of the opposite lobes determined {\sl independently} of each other suggest that
their ratios are between $\sim$1.1 to $\sim$1.4. Demanding similar values of the jet power and
the radio core density for the same GRS, we look for a {\sl self-consistent} solution for
the opposite lobes, which results in different density profiles along them found by the fit.
A comparison of the dynamical and spectral ages shows that their ratio is between $\sim$1 and
$\sim$5, i.e. is similar to that found for smaller radio galaxies. Two causes of this effect
are pointed out.

\end{abstract}


\section{Introduction}   
The dynamical ages of the opposite lobes of 10 selected giant radio sources
are estimated using the DYNAGE algorithm of Machalski et al. (2007) and
compared with their spectral ages determined and studied by Jamrozy et al.
(2008). The DYNAGE algorithm is an extension of the analytical model for the
evolution of FRII type radio sources combining the dynamical model of Kaiser
\& Alexander (1997) with the model for expected radio emission from a source
under the influence of energy loss processes published by Kaiser, Dennett-Thrope, \& Alexander 1997;
hereafter referred to as KDA). This algorithm allows to determine the values
of four of the model parameters, i.e. the jet power, $Q_{\rm jet}$, central
core density, $\rho_{0}$, energy distribution of the relativistic particles
injected into the lobe via acceleration processes and described by the spectral
index $\alpha_{\rm inj}$, and the lobe's age, $t$. The determination of their
values is made possible by the fit to the observational parameters of a lobe:
its length, $D$, axial ratio (geometry), $R_{\rm T}$, normalization and slope
of its radio spectrum, i.e. the radio luminosity, $P_{\nu,i}$, at a number of
observing frequencies $\nu$=1,2,3... The values of other free parameters of
the model have to be assumed.

\section{Dynamical Model and the Age Solutions}

Three basic relations of the model are:

\begin{equation}
\rho(r)=\rho_{0}(r/a_{0})^{-\beta}\hspace{7mm} {\rm for}\hspace{2mm} r\geq a_{0},
\end{equation}

\begin{equation}
L_{\rm jet}(t)={\rm const}\left(Q_{\rm jet}/\rho_{0}a_{0}^{\beta}\right)
^{1/(5-\beta)}t^{3/(5-\beta)}\hspace{10mm}{\rm and,}
\end{equation}

\begin{equation}
P_{\nu}=\frac{1}{6\pi}\int_{(t_{min})}^{t}\sigma_{\rm T}c\,u_{\rm B}Q_{\rm jet}
\frac{\gamma^{3}}{\nu}n(\gamma, t_{\rm i})V(t_{\rm i})dt_{\rm i}
\end{equation}

\subsection{Independent Age Solution {\rm (3 steps)}}

1) Assuming the values of remaining free parameters of the model (cf. Machalski et al. 2009) --
for given values of $p(\alpha_{\rm inj})$ and $t$ equating
$$L_{\rm jet}\overbrace{=}^{[Equation\,2]}D_{\rm obs}/\sin\Theta\hspace{15mm}$$
and
$$P_{\nu}
\overbrace{=}^{[Equation\,3]}P_{\nu,obs},$$

the DYNAGE fits the values of $Q_{\rm jet}(\alpha_{\rm inj},t,P_{\nu})$ and 
$\rho_{0}(\alpha_{\rm inj},t,P_{\nu})$.

\noindent 2) Performing step 1) for all $P_{\nu}$ chosen to represent the lobe's spectrum, we have
$t(\alpha_{\rm inj})$, $Q_{\rm jet}(\alpha_{\rm inj})$, and
$\rho_{0}(\alpha_{\rm inj})$ (Figure\,1, { left}).

\noindent 3) The steps 1) and 2) are repeated for a number of $\alpha_{\rm inj}$ values searching for
its value which provides a minimum of the kinetic energy delivered to the lobe at the age $t$,
i.e. $Q_{\rm jet}\times t$ (Figure\,1, {right}). This, in turn,
provides the final solutions of $t$, $Q_{\rm jet}$, and $\rho_{0}$, as well as of their derivative
parameters (see Table~4 in Machalski et al. 2007)

\begin{figure}
\plottwo{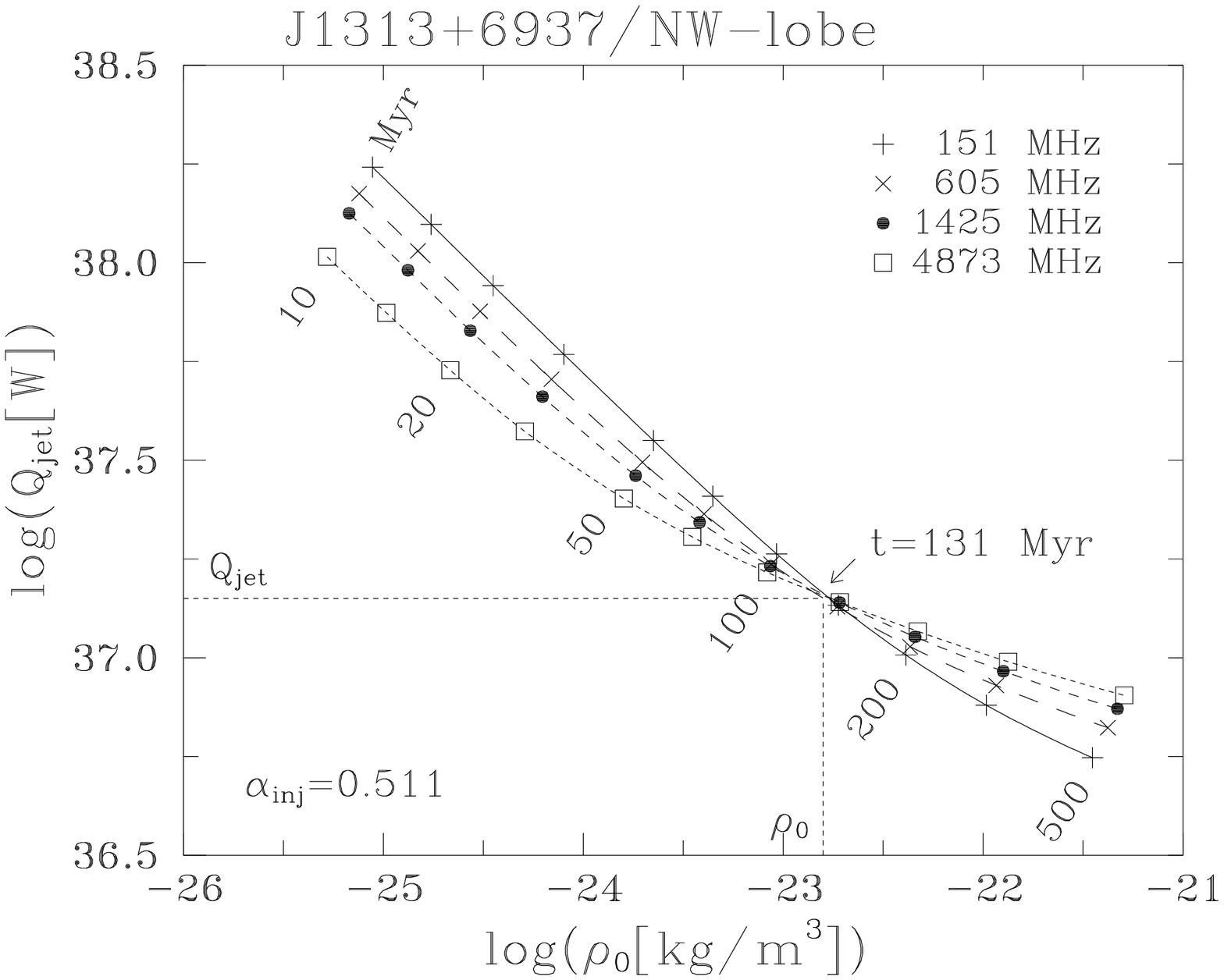}{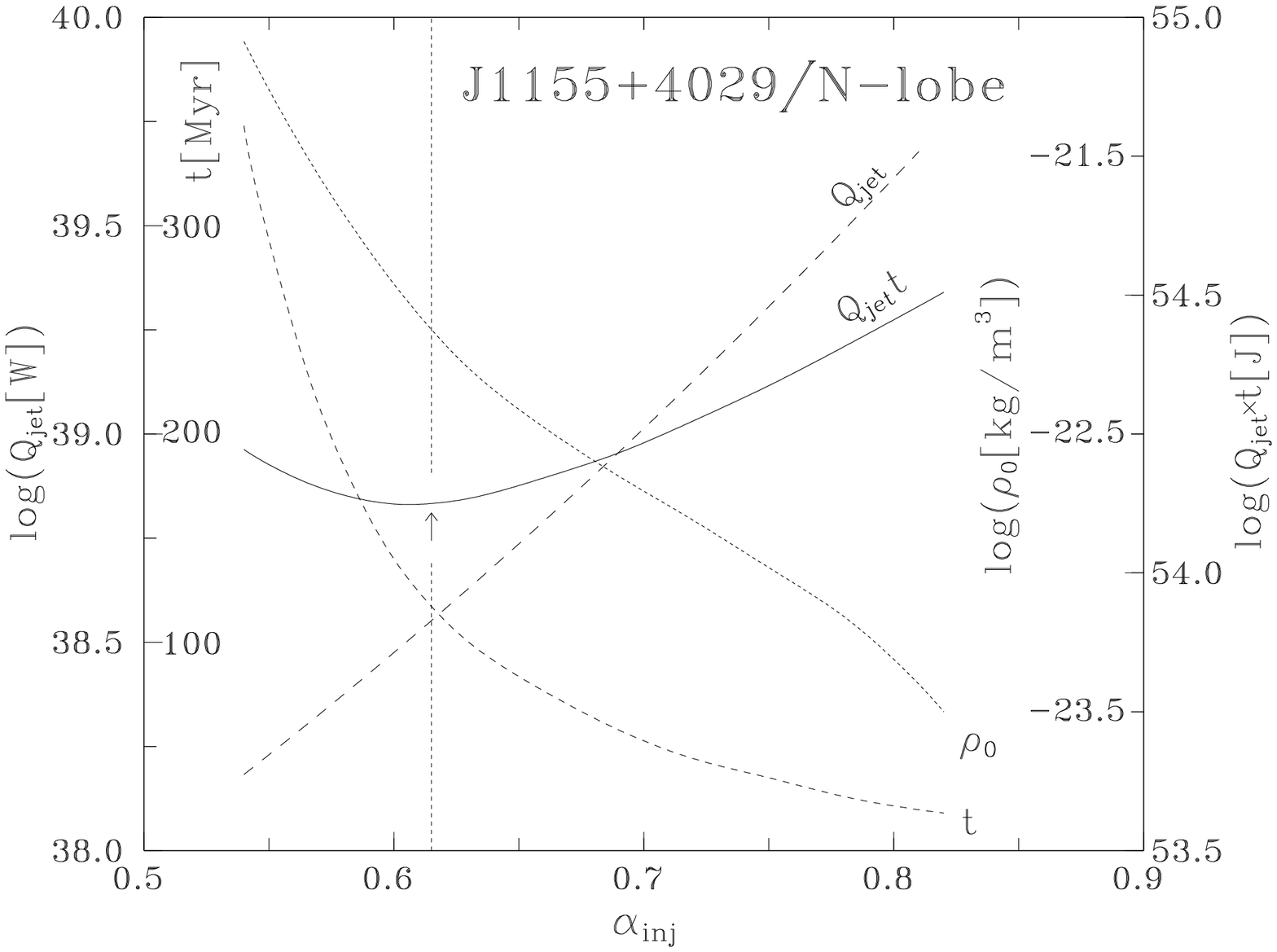}
\caption{{Left:\/} A set of $Q_{\rm jet}(t)$ and $\rho_{0}(t)$ solutions at the four
observing frequencies for a lobe of J1313+6937.
 {Right:\/} The dependence of $Q_{\rm jet}$, $t$, $Q_{\rm jet}\times t$, and $\rho_{0}$
on $\alpha_{\rm inj}$ for a lobe of J1155+4029.}
 \end{figure}

{\bf Objections:} Independent age solutions give {\sl different ages} of the opposite lobes
which is shown in Figure\,2 {(left)}. The differences are larger than expected kinematic age
differences due to a projection of the jet's axis.

\noindent
From Equation\,(2) we have

\begin{equation}
t=\left(\frac{D}{c_{1}}\right)^{(5-\beta)/3}\left(\frac{\rho_{0}a_{0}^{\beta}}{Q_{\rm jet}}\right)^{1/3}.
\end{equation}

\noindent
For the same source (consisting of two lobes) we note that

-- larger $D$ suggests older $t$, and calculations show that (usually)

-- larger $D$ accompanies a higher value of $\rho_{0}$ than that for shorter $D$.

\noindent
Thus, substitution of $\langle\rho_{0}\rangle$ into Equation\,(4) and a change of $\beta$ can lower $t$.
However, this does not satisfy Equation\,(3) where the lobe's radiation is controlled by $Q_{\rm jet}$
and $\alpha_{\rm inj}$. Therefore we look for self-consistent age solution.

\subsection{Self-Consistent Age Solution}

\noindent
Let $\rho(r=D)\equiv\rho_{\rm a}$; thus from Equation\,(1)
$$\rho_{0}\left(\frac{D}{a_{0}}\right)^{-\beta}=\langle\rho_{0}\rangle\left(\frac{D}{a_{0}}\right)
^{-\beta_{\rm s.c.}}\hspace{10mm}{\rm we\hspace{2mm} have}\hspace{5mm}\beta_{\rm s.c.}$$

\noindent
Substitution of $\beta_{\rm s.c.}$ into Equation\,(4) gives the `self-consistent' age

\begin{equation}
t_{\rm s.c.}=\left(\frac{D}{c_{1}}\right)^{(5-\beta_{\rm s.c.})/3}
\left(\frac{\langle\rho_{0}\rangle a_{0}^{\beta_{\rm s.c.}}}{\langle Q{\rm jet}\rangle}\right)^{1/3}
\end{equation}

\noindent
The `self-consistent' ages of the opposite lobes calculated from Equation\,(5) are plotted vs
$\beta_{\rm s.c.}$ in Figure\,2 {(right)}.

 \begin{figure}
\plottwo{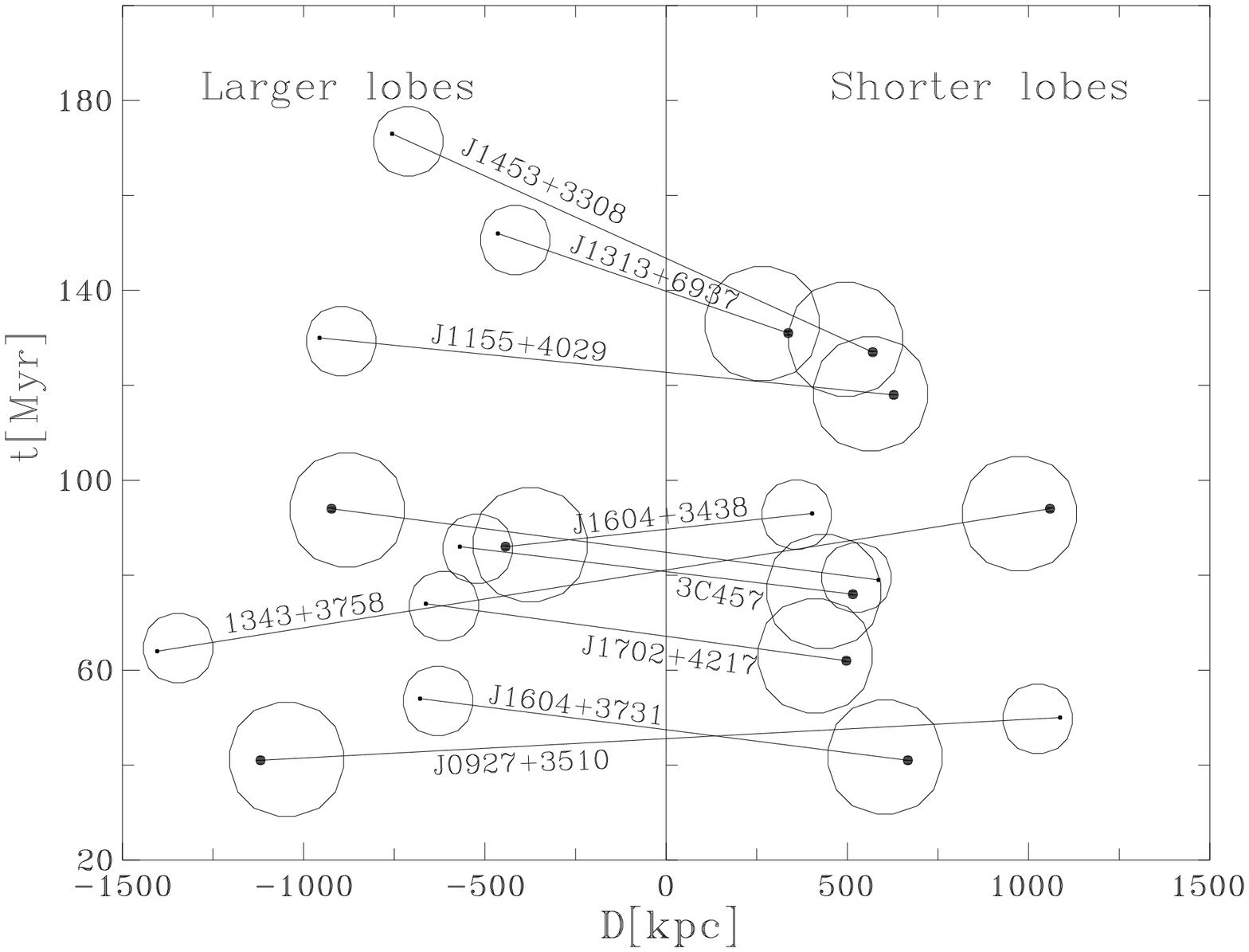}{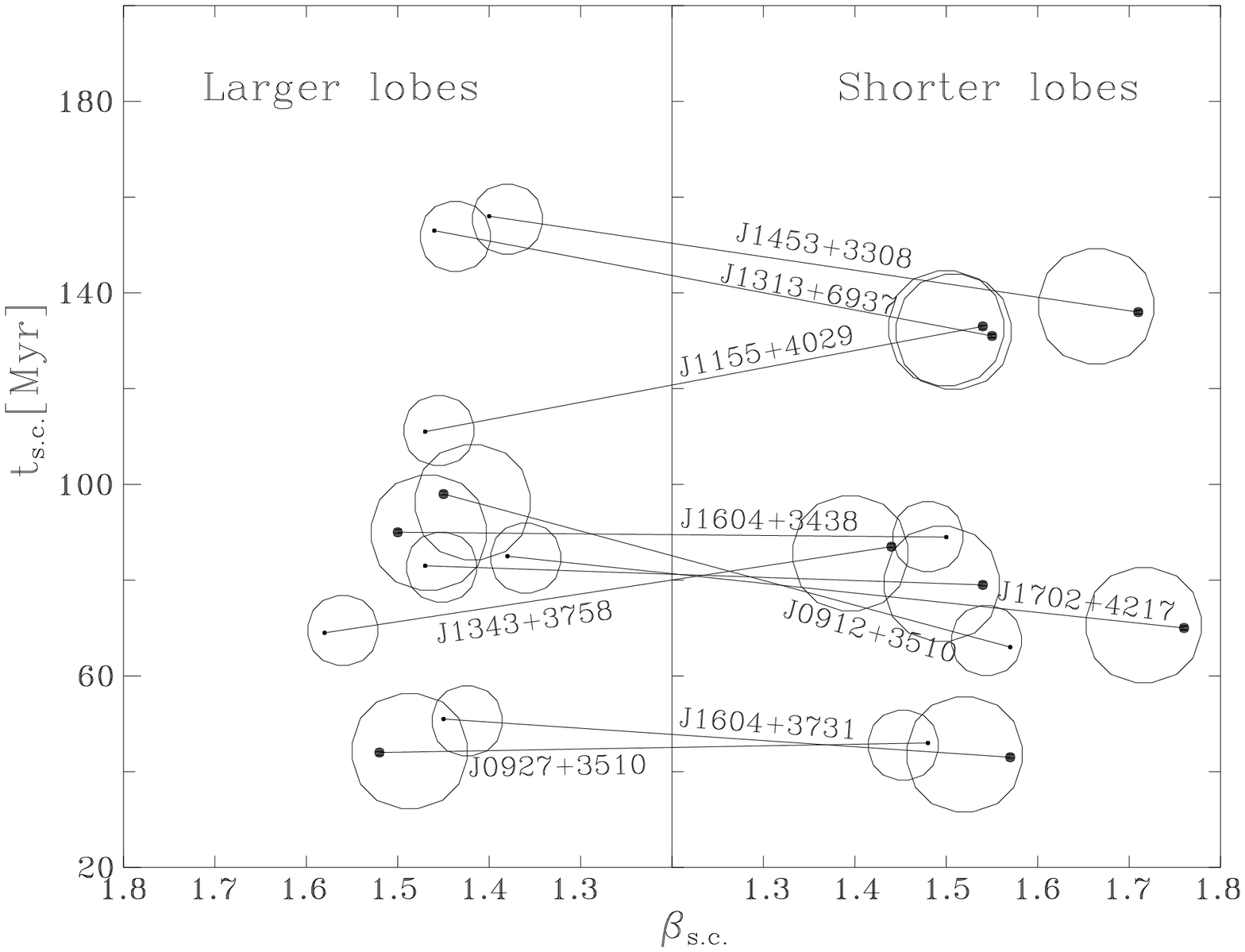}
 \caption{{Left:\/} Schematic display of the `independent solution' of age, $t$ vs length
of the lobes.
 {Right:\/} Similar display of the `self-consistent' solution (the same $Q_{\rm jet}$
and $\rho_{0}$ forced for both lobes) of age, $t_{\rm s.c.}$, vs exponent of the external density
profile. In both diagrams longer lobes are on the left side, while shorter ones -- on the right
side of the diagram.  Larger circles indicate a brighter lobe; smaller circles -- a fainter one.}
 \end{figure}

\begin{figure}
\noindent\hfil\psfig{figure=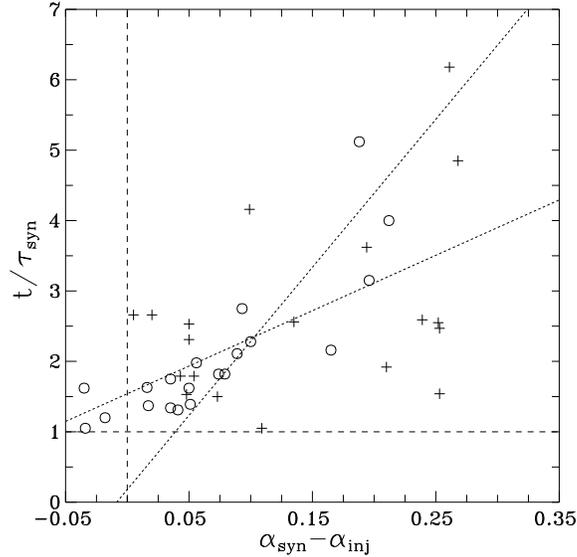,width=3in}
\caption{Ratio of the dynamical age of the lobe, $t$, and their
synchrotron age, $\tau_{\rm syn}$, vs the difference $\alpha_{\rm syn} -
\alpha_{\rm inj}$. Circles indicate the ratio of $t$, and $\tau_{\rm
CI}$, while crosses indicate the ratio of $t$, and $\tau_{\rm JP}$. The
value of $\tau_{\rm CI}$ and $\tau_{\rm JP}$ are given in Jamrozy et al.
(2008). The two dotted lines show linear regression lines on each of the
two coordinates.}
 \end{figure}

\section{Discussion of the Results}
\begin{enumerate}

\item  In 7 of 10 GRSs the brighter lobe is found to be younger than the opposite, fainter one, but 

\item  dynamical ages of opposite lobes differ between themselves more than it is expected due to
the kinematic age difference related to a projection of the jet's axis toward the observer.

\item  The apparent asymmetries in the lobes' length, their luminosity, and age are likely to be  due to
different propagation conditions of the jets through an external medium.

\item  Assuming different exponents $\beta$ in Equation\,(1) in opposite jet's directions, the age
difference usually decreases, however still remaining larger than that due to the kinematic effect.  

\item  Expected exponents $\beta_{\rm s.c.}$ for the shorther lobes are higher that those for the
larger ones.

\item  Differences of age $t_{\rm s.c.}$ can be levelled by a departure from the equipartition
between the energy densities of magnetic fields and relativistic particles. An increase of
$r\equiv u_{\rm B}/u_{\rm c}$ results in increase of the lobe's age; however it is connected
with a corresponding decrease of the jet power. A decrease of $r$ acts inversely.

\item  A ratio of the dynamical age to the spectral age is between 1 and 5. This is caused by
(i) a difference between the injection spectral indices determined using the DYNAGE algorithm
in the dynamical analysis and the SYNAGE algorithm of Murgia (1996) in the spectral-ageing
analysis (shown in Figure\,3), and (ii) a different influence of the axial ratio of the lobes
in the estimation of the dynamical age and the spectral age. 
\end{enumerate}

In Machalski, Jamrozy, \& Saikia  (2009) arguments are given that DYNAGE can better take account of radiative
effects at low frequencies than SYNAGE, and the age solutions should be better than those found
with the classical spectral-ageing analysis because the expansion parameters are connected to
actual geometry of the lobes for each specific GRS. The DYNAGE algorithm is especially effective
for sources at high redshifts for which an intrinsic spectral curvature is shifted to low frequencies.


\begin{thebibliography}{}

\bibitem[]{}Jamrozy, M., Konar, C., Machalski, J., \& Saikia, D. J. 2008, MNRAS, 385, 1286
\bibitem[]{}Kaiser, C. R.,  \& Alexander, P. 1997, MNRAS, 286, 215
\bibitem[]{}Kaiser, C. R., Dennett-Thorpe, J., \& Alexander, P. 1997, MNRAS, 292, 723 (KDA)
\bibitem[]{}Machalski, J., Chy\.{z}y, K.T., Stawarz, L., \& Koziel, D. 2007, A\&A, 462, 43
\bibitem[]{}Machalski, J., Jamrozy, M., \& Saikia, D. J., 2009, MNRAS, 395, 812
\bibitem[]{}Murgia, M. 1996, Ph.D Laurea Thesis, University of Bologna

\end{thebibliography}
\end{document}